\documentstyle[12pt]{article}
\begin{document}
\begin{center}
{\bf COVARIANT QUANTUM DYNAMICAL SEMIGROUPS:\\ UNBOUNDED GENERATORS}

\vskip30pt
{\bf A.S.Holevo}

Steklov Mathematical Institute,\\ Vavilova 42, 117966 Moscow, Russia\\
E-mail: holevo@class.mi.ras.ru
\end{center}
\vskip30pt
{\bf Summary}. {\sl A survey of probabilistic approaches to quantum dynamical
semigroups with unbounded generators is given. An emphasis is made upon
recent advances in the structural theory of covariant Markovian master
equations. As an
example, a complete characterizations of the Galilean covariant irreversible
quantum Markovian evolutions is given in terms of the corresponding quantum
master and Langevin equations. Important topics for
future investigation are outlined.}

\section{Introduction}
Quantum dynamical semigroups are a noncommutative analog of (sub-) Markov
semigroups in classical probability: while the latter are semigroups of maps
in functional spaces, the former are semigroups of maps in operator algebras,
having certain properties of positivity and normalization. In quantum
statistical mechanics dynamical semigroups arise when one considers weak or
singular coupling or low density limits for open quantum system interacting
with surrounding, allowing to neglect the memory effects of the interaction
\cite{spohn}.  These semigroups satisfy differential equations that are
noncommutative generalization of the Fokker-Planck or Chapman-Kolmogorov
equations and represent the general solution of the Cauchy problem for such
equations.

 Let ${\cal B}({\cal H})$ be the algebra of all bounded operators in a
Hilbert space $\cal H$. We denote by $I$ the unit operator in $\cal H$, and
by  Id  the identity map of ${\cal B}({\cal H})$.  Since ${\cal B}({\cal H})$
is the dual Banach space of the space ${\cal T}({\cal H})$ of trace-class
operators, it is supplied with the {\sl weak$^*$ topology}. On norm-bounded
sets this topology coincides with the weak operator topology (see e.g.
\cite{brat} ).  A bounded map $\Phi $ of ${\cal B}({\cal H} )$ into itself is
{\sl completely positive} (c. p.) if $$
\sum \limits_{i,j}(\psi_{i} |\Phi [X_{i}^{*}X_{j}]\psi_{j}) \geq 0 \eqno(1)$$
for any finite sets $\{ \psi_j \} \in {\cal H} , \{ X_j \} \in {\cal B}({\cal
H}).$ According to Stinespring's theorem adapted to the case of ${\cal
B}({\cal H})$ (see \cite{kraus}), a generic weak$^*$-continuous c. p. map has
the representation $$\Phi [X] = L^* (X\otimes I_0 )L, \eqno(2)$$ where $L$ is
a bounded operator from $\cal H$ to ${\cal H}\otimes{\cal H}_0$ and $I_0$ is
the unit operator in an auxiliary Hilbert space ${\cal H}_0$.  By using this
representation, it is possible to show that such maps satisfying additional
normalization condition $\Phi [I] = I$ represent irreversible evolutions of
the open quantum system interacting via unitary operator with auxiliary
system in a fixed initial state (see \cite{kraus}, \cite{lewis}).

If $I_0 = \int |\psi_x ><\psi_x |\mu (dx)$ is a resolution of identity in
${\cal H}_0$, then (2) implies $$\Phi [X] = \int L(x)^* XL(x)\mu (x),$$ where
$L(x)=(I\otimes\psi_x )^* L$, and $I\otimes\psi_x$ is the operator
from $\cal H$ to ${\cal
H}\otimes{\cal H}_0$, mapping $\psi$ into $\psi\otimes\psi_x$. In particular,
if the measure $\mu (dx)$ is discrete, we obtain the familiar
representation of a normal c. p. map in ${\cal B}({\cal H} )$.

By a {\sl dynamical semigroup} in ${\cal B}({\cal H})$ (or {\sl quantum
dynamical semigroup}) we shall call a semigroup $\Phi_t ;~ t\geq 0,$ of
weak$^*$ continuous completely positive maps in ${\cal B}({\cal H})$,
satisfying $\Phi_0 = \mbox{Id}$, and $\Phi_t [I]\leq I$.  Moreover, for any
$X$ the function $t \rightarrow \Phi_t [X]$ is required to  be
weak$^*$-continuous.  $\Phi_t$ is called {\sl unital} if $\Phi_t [I] = I$.

In the case of finite-dimensional $\cal H$ the weak$^*$ continuity is
equivalent to the norm continuity; every quantum dynamical semigroup then has
the form $\Phi_t = \mbox{exp}t{\cal L}$, where $\cal L$ is the {\sl
generator} of the semigroup. The generator is {\sl conditionally completely
positive} map, which means that inequality of the type (1) holds provided
$\sum \limits_j X_j \psi_j = 0,$ and satisfies the normalization condition
${\cal L}[I]\leq 0$ (or ${\cal L}[I] = 0$ for unital semigroup). The
semigroup is the unique solution of the {\sl backward} and the {\sl forward
Markovian master equations}  \break (M. m. e.) $$\frac{d}{dt}\Phi_t = {\cal
L} \circ\Phi_t ;\qquad\frac{d}{dt}\Phi_t = \Phi_t\circ{\cal L},\eqno(3)$$
satisfying $ \Phi_0 = \mbox{Id}$.

The conditional complete positivity of $\cal L$ is equivalent to the {\sl
standard representation}  $${\cal L}[X] =  \Phi [X] - K^* X - XK,  \eqno(4)$$
where $ \Phi $ is a c. p. map of the form (2), and the normalization
condition is equivalent to $L^* L\leq K^* + K$ (with equality for unital
semigroups). Similar results hold for {\sl norm-continuous} semigroups in
infinite-dimensional $\cal H$ with all operators in question being {\sl
bounded}. The representation (4) for this case was established by Lindblad
\cite{lindblad}, and independently  an equivalent representation was obtained
by Gorini, Kossakowski and Sudarshan \cite{gorini} for dim${\cal H}< \infty$.
A physical interpretation for the standard representation can be seen from
the Dyson expansion of the solution of the forward~~ M. m. e.  $$\Phi_t =
{\hat \Phi}_t + \sum_{n=1}^{\infty}\int...\int_{0\leq t_1 \leq ...\leq t_n
\leq t}{\hat \Phi}_{t_1} \circ\Phi\circ{\hat \Phi}_{t_2-t_1}
...\Phi\circ{\hat \Phi}_{t-t_{n}} dt_1 ...dt_{n} \eqno(5)$$ described as
sequence of ``spontaneous jumps'' of the magnitude $\Phi$ occuring at times
$t_1 \leq ...  \leq t_n$ on the background of the ``relaxing evolution''
given by the semigroup ${\hat \Phi}_t [X] = \mbox{e}^{-K^* t}X\mbox{e}^{-K
t}$.

Let $g\rightarrow V_g$ be a unitary representation of a group $G$ in $\cal
H$.  The dynamical semigroup $ \Phi_t$ is called {\sl covariant} if $$ \Phi_t
[V_g^* XV_g ] = V_g^*  \Phi_t [X] V_g .   \eqno(6)$$ The property of
covariance reflects presence of certain symmetries in the interacting open
quantum system and is important for applications. For example, covariance
with respect to various subgroups of the orthogonal group is characteristic
for Bloch type equations and is relevant for optical or magnetic resonance
spectroscopy \cite{alicki} or theory of anisotropic relaxation of spin
systems \cite{artem}. The combination of covariance and complete positivity
imposes strong restrictions on the generator, in some cases defining
practically uniquely the form of the corresponding M. m. e..  It can be shown
\cite{hol1} that the generator of a covariant norm continuous semigroup
admits the representation (4) in which  $\Phi$ is covariant and $K$ commutes
with $V_g$ (provided $G$ is amenable). This in general is not true for
non-norm continuous case, see Section 4.

Already in the mid-seventies when the structure of norm-continuous dynamical
semigroups was well understood, it became clear that the  non-norm continuous
case, while being interesting from both physical and mathematical points of
views, poses difficult problems. The generator of such a semigroup may be
unbounded with domain not even necessarily being a * - algebra.  Therefore it
was difficult to formulate a generalization of conditional complete
positivity (or equivalent property) useful enough to obtain a kind of
standard representation for $\cal L$ with all its important consequences.
The very formulation of the standard representation needed clarification and
it was not obvious whether there are ``non-standard'' unbounded generators.
Among very few papers on the subject, Davies \cite{dav1} established a
standard representation for semigroups on ${\cal B}({\cal H})$ having
invariant pure state. Bratteli et al. \cite{brat} studied semigroups on
rather general C$^*$-algebra covariant with respect to a compact Abelian
group and satisfying rather strong restriction that $\cal L$ vanishes on the
fixed point subalgebra, and showed a kind of Levy-Khinchin formula for $\cal
L$.  There were few papers on quasi-free dynamical semigroups on CCR algebra,
generators of which are certainly unbounded and standard, e. g. \cite{emch},
\cite{vanheuv}. Unbounded generators arise when the semigroup is covariant
with respect to a non-compact symmetry group (such as translations or Galilei
group, \cite{lanz}). While enormous attention was paid to the study of
reversible evolutions generated by Schr\"odinger operators, much less is
known about their irreversible Markovian counterparts.

A substantial progress in this direction was achieved in the past few years
by making use of profound analogies from the classical theory of Markov
semigroups, as developed by Feller, Dynkin, Ito and McKeane, see e. g.
\cite{feller}, \cite{ito}, or by direct use of classical probabilistic
methods. This development concerns the following topics:  \begin{itemize}
\item The minimal dynamical semigroup \cite{dav2}, \cite{chebot}, \cite{chf},
 \cite{hol2}, \cite{sinha}. Existence and uniqueness of solutions of
Markovian master equations \cite{hol3}, \cite{hol9}; \item The structure of
(covariant) Markovian master equations \cite{hol2}, \cite{hol6}, \cite{hol3},
\cite{hol7}; \item Noncommutative excessive functions and arrival times
\cite{bhat}, \cite{hol4}.  Non-standard generators \cite{hol4}, \cite{hol5};
\item Stochastic representations and hyperdissipativity \cite{hol9}.
Relations to continuous measurement processes and nonlinear stochastic
Schr\"odinger equations \cite{gat}, \cite{hol9}, \cite{kol}. Dilations to
quantum Langevin equations \cite{hol7}.  \end{itemize}

In what follows we shall concentrate on the second topic, restricting to
brief comments concerning other topics; further details can be found in the
references given above.  \section{The quantum Markovian master equations} The
starting point of our approach, just as in the classical probability theory,
is not a semigroup itself, but the differential equation it satisfies.  This
is also more natural for physical applications. A quantum M.  m. e. must be
an equation for matrix elements of the semigroup; thus we assume that there
is a dense domain $\cal D \subset \cal H$, such that the following derivative
$$\frac{d}{dt}\left.<\psi |\Phi_t [X]\phi > \right|_{t=0} = {\cal L}(\psi ;
X; \phi ) \eqno(7)$$ exists for $\phi, \psi \in \cal D$, $X\in {\cal B}({\cal
H}) $.  The form ${\cal L}(\psi ; X; \phi )$ is called {\sl form-generator}.
It can be characterized by a number of nice properties including conditional
complete positivity \cite{hol2}, \cite{hol3}. These properties turn out to be
equivalent to the {\sl standard representation} $${\cal L} (\psi ;X ;\phi ) =
<L \psi |(X\otimes I_0 )L \phi > - <K\psi |X\phi > - <\psi |XK\phi >, \eqno
(8)$$ where $L, K$ are (unbounded) operators defined on $\cal D$ and
satisfying the {\sl dissipativity condition} $$ ||L\psi ||^2 \leq~ 2\mbox{
Re}<\psi |K\psi >,~~~\psi \in \cal D. $$ In particular, $K$ is accretive:
$\mbox{ Re}<\psi |K\psi >\geq 0, \psi \in \cal D.$ The (backward) M. m. e.
takes the form $$\frac{d}{dt}<\psi |\Phi_t [X]\phi > = {\cal L}(\psi ; \Phi_t
[X]; \phi );~~ \phi ,\psi \in \cal D.   \eqno (9)$$

The relation between the form-generator and the generator resembles relation
between a formal differential operator and its closed extensions determined
by certain boundary conditions. To see this let $\Psi_t = (\Phi_t )_*$ be the
strongly continuous {\sl preadjoint semigroup}  in ${\cal T}({\cal H})$, such
that ${\Psi_t}^* = \Phi_t$.  Denoting its generator ${\cal L}_*$, one has
$${\cal L} (\psi ;X ;\phi ) = \mbox{ Tr} {\cal L}_* [|\phi ><\psi |]X$$ for
$\phi ,\psi\in \cal D$, $X\in{\cal B}({\cal H})$. The assumption that the
derivative (7) exists for all $X$ is equivalent to dom${\cal L}_* \supset
{\sf D}$, where $$\mbox{\sf D} =\mbox{ lin}\{|\phi ><\psi |: ~~\phi, \psi \in
\cal D \} \eqno(10)$$ is a dense domain in  ${\cal T}({\cal H})$. The M. m.
e. (9) takes the form $$\frac{d}{dt}\mbox{ Tr}\rho \Phi_t [X] = \mbox{
Tr}{\cal L}_* [\rho ]\Phi_t [X], \qquad\rho\in\mbox{\sf D}, X\in {\cal
B}({\cal H}). \eqno (11)$$

If \mbox{\sf D} is a core for ${\cal L}_*$, then this equation determines
$\Phi_t$ uniquely, otherwise it may have non-unique solution. Under the
condition that the closure of $K$ is maximal accretive, one can show that
there exists a dynamical semigroup ${\Phi_t}^{\infty}$ giving the {\sl
minimal solution} of the equation (9) in the sense that for any other
solution $\Phi_t$ the difference $\Phi_t - {\Phi_t}^{\infty}$ is completely
positive. Of special interest is the case of a {\sl unital} generator,
satisfying  $\mbox{ Tr}{\cal L}_* [\rho] \equiv 0, \rho\in${\sf D}, or $$||L
\psi ||^2 = 2\mbox{ Re} <\psi |K\psi >,~~\psi \in \cal D.$$  In general
${\Phi_t}^{\infty}$ may not be unital; however if it is, then
${\Phi_t}^{\infty}$ is the unique solution of (9).

The method of construction of the minimal dynamical semigroup developed in
\cite{dav2} for resolvents, in \cite{chebot}, \cite{chf} for associated
integral equation, and in \cite{hol2}, \cite{hol3} for the backward M. m. e.,
is the noncommutative extension of the Feller's method \cite{feller}. It is
based on a standard representation, i. e. on a decomposition of the relevant
object into completely positive and relaxing parts. The starting point is the
relaxing semigroup ${\hat \Phi}_t [X ] = \mbox{e}^{-{\bar K}^*
t}X\mbox{e}^{-{\bar K}t}$ providing the unique solution of the equation (9)
with $${\cal L} (\psi ;X ;\phi ) = - <K\psi |X\phi > - <\psi |XK\phi >,$$
which is then perturbed with the completely positive form $<L \psi |(X\otimes
I_0 )L \phi >$ introducing spontaneous jumps on the background of the
relaxing evolution ${\hat \Phi}_t$.  It may be viewed upon as a
generalization of the expansion (5) to the case of unbounded but completely
positive perturbations.  Just as in the classical case, ``explosion'' may
occur if the infinite number of jumps happens during finite interval and the
process reaches ``boundary'' in a finite time (this can never happen for a
bounded generator).  If $\Phi_t^{\infty}$ is not unital, then there is a
positive probability of explosion, and additional ``boundary conditions'' are
required to specify the solution, which amounts to certain maximal extension
of ${\cal L}_*$ from {\sf D}.

 Under the additional assumption that operator ${L}^*$ satisfies $$\sum_j
||L^* (\psi\otimes e_j )||^2 < \infty,~~~\psi \in {\cal D}^* , $$ where
$\{e_j\}$ is an orthonormal basis in ${\cal H}_0$, and  ${\cal D}^*
\subset\mbox{dom}K^*$ is a dense domain in ${\cal H}_0$, one can write also
the {\sl forward} Markovian master equation for the preadjoint semigroup
$\Psi_t$: $$\frac{d}{dt} <\phi |\Psi_t [\rho ]\psi > = {\cal L}_* (\phi;
\Psi_t [\rho ];\psi );~~ \phi ,\psi \in {\cal D}^*, \eqno (12)$$ where
$${\cal L}_* (\phi ;\rho ;\psi ) = \mbox{Tr}\rho{\cal L}[|\psi><\phi|]$$ $$ =
\sum_j < L^* (\phi\otimes e_j )|\rho L^* (\psi \otimes e_j )> - <K^* \phi
|\rho \psi > - <\phi |\rho K^* \psi >, $$ and $\cal L$ is the generator of
$\Phi_t$ defined on {\sf D}$^* = \{ |\psi><\phi| :  \phi, \psi \in {\cal
D}^*\}.$ Assuming $K^*$ to be maximal accretive one can prove that
${\Psi}_t^{\infty} = ({\Phi_t}^{\infty})_*$ is the minimal solution of the
forward equation \cite{hol7}, \cite{hol9}.  However in general the forward
and the backward equations are no longer equivalent. Thus the situation is
similar to that for the Kolmogorov-Feller differential equations in the
theory of Markov processes \cite{feller}.

Going back to the problem of standard representation, we can make the
following remarks. The fact that a form-generator has the standard
representation (8) implies the possibility of decomposing the generator
${\cal L}_*$ into completely positive and relaxing parts only on the subspace
{\sf D} which need not be a core for ${\cal L}_*$. If explosion occurs, these
two parts need not be separately extendable onto a core for ${\cal L}_*$.  On
the other hand, generators of different dynamical semigroups restricted to
{\sf D} can give rise to one and the same standard expression (8). One may
formalize the notion of standard representation by saying that a dynamical
semigroup is {\sl standard} if it can be constructed as the minimal semigroup
for some M. m. e., that is by a completely positive perturbation of a
relaxing semigroup. In \cite{hol4} a possible noncommutative extension of
``boundary conditions'' for conservative form-generator was proposed as very
singular completely positive perturbations vanishing on the dense domain {\sf
D}. By using such a perturbation the author gave a construction of
non-standard dynamical semigroup on ${\cal B}({\cal H})$ \cite{hol4},
\cite{hol5}.
\section{An example} Let $\xi_t, t\geq 0$ be stochastic process
with stationary independent increments \cite{feller}. Roughly speaking, the
(generalized) time derivative of $\xi_t$ is a continuous analog of a sequence
of independent identically distributed random variables, that is a classical
``noise'' process.  One of the beautiful results of probability theory is the
Levy-Khinchin formula describing the possible form of the characteristic
function of such process:  $${\sf M}\mbox{exp}i\lambda\xi_t =
\mbox{exp}t[i\beta\lambda - \frac{\alpha}{2}\lambda^2
+\int_{0<|y|}(\mbox{e}^{iy\lambda}-1-iy\lambda 1_h(y))\mu (dy)], \eqno(13)$$
where  $\beta$ is real number, $\alpha\geq0$, $h$ is arbitrary but fixed
positive number, $1_h(y)$ is the indicator of the set $|y|\leq h$, and $\mu
(dy)$ is a positive measure on the set $\mbox{\bf R}\setminus\{0\}$,
satisfying the condition $$\int_{0<|y|}[y^2 1_h(y)+(1-1_h(y))]\mu(dy)<\infty.
\eqno(14)$$ In (13) the term $ i\beta\lambda -\frac{\alpha}{2}\lambda^2 $
corresponds to the Gaussian component of the process $\xi_t$, which is a
continuous process.  If in the integral term we take $\mu (dy ) =\mu~ \delta
(y-y_0 )dy$ with $\mu >0$, then for $|y_0 |>h$ we obtain logarithm of the
characteristic function of the Poisson process with the jumps of the
magnitude $y_0$. Therefore for arbitrary measure $\mu (dy)$ the integral
$\int_{h<|y|}(\mbox{e}^{iy\lambda}-1)\mu (dy)$ describes the mixture of
independent Poisson processes with various magnitudes $y,~ |y|>h.$ It
corresponds to the discontinuous (pure jump) component of the process $\xi_t$
(with magnitudes of jumps $|y|>h$).  The value of $h$ is arbitrary but fixed,
so the name "big jumps" is only conventional.  The term related to "small
jumps" (of magnitudes $|y|\leq h$) corresponds to the situation when
infinitely many small jumps can accumulate during finite time, and one must
include portions of linear drift between jumps in order that the total
increment  will remain finite. The process $\xi_t$ itself can be decomposed
into three components -- continuous Gaussian, Poisson ``big jumps'' and
``small jumps'', according to Ito's formula (see e. g.  \cite{stoch}):
$$d\xi_t = \beta dt+\sqrt{\alpha}
dW_t+\int_{h<|y|}y\Pi(dy~dt)+\int_{0<|y|\leq h} {\tilde \Pi}(dy~dt),
\eqno(15)$$ where $W_t$ is the standard Wiener process, $\Pi(dy~dt)$ is the
Poisson random measure on ${\bf R}^2$ with the compensator $\mu (dy) dt$, so
that $${\sf M}dW_t = 0,\quad {\sf M}\Pi(dy~dt) = \mu(dy)~dt, \eqno(16)$$ and
${\tilde \Pi}(dy~dt) ={\Pi}(dy~dt)- \mu(dy)~dt$ is the compensated random
measure. Note that $\Pi([y_1, y_2], [t_1, t_2])$ is just the number of jumps
of the process $\xi_t$ on the time interval $[t_1, t_2]$, which have
magnitudes $y\in[y_1, y_2]$.

Now consider the Hilbert space ${\cal H}=L^2({\bf R})$, and let $Q=x, P=
i^{-1}\frac{d}{dx}$ be, respectively, the self-adjoint position and momentum
operators for one-dimensional quantum system, so that $V_y=\mbox{exp}(iyQ),
y\in{\bf R},$ and $U_x=\mbox{exp}(-ixP), x\in{\bf R},$ are the unitary groups
in $\cal H$ satisfying the Weyl canonical commutation relation (CCR): $$U_x
V_y = \mbox{exp}(-ixy) V_y U_x. \eqno(17)$$ Defining $$\Phi_t [X] = {\sf
M}U_{\xi_t}^* X U_{\xi_t},    \quad t\geq 0, \eqno(18)$$ one easily sees that
$\Phi_t$ is a unital dynamical semigroup in $\cal H$.  Indeed, operators
$\Phi_t$ are manifestly completely positive; the semigroup property follows
from the fact that $\xi_t$ has stationary independent increments; the
weak$^*$ continuity properties follow from the continuity properties of $U_x$
and of the expectation. The semigroup (18) represents the dynamics of quantum
system in $\cal H$ interacting with the classical noise via unitary operators
$\mbox{exp}(-i\xi_t P)$, averaged with respect to the distribution of the
noise. To find the generator of this semigroup, one can use the Ito formula
for $\mbox{exp}(i\xi_t P)$ (cf.  \cite{hol8}): $$d\mbox{exp}(i\xi_t
P)=\mbox{exp}(i\xi_t P) \{[i\beta P - \frac{\alpha}{2}P^2 +\int_{0<|y|\leq
h}(\mbox{exp}(iyP)-1-iyP) \mu(dy)]dt$$ $$+ i\sqrt{\alpha}P dW_t +
\int_{h<|y|}[\mbox{exp}(iyP)-1]\Pi(dy~dt) + \int_{0<|y|\leq
h}[\mbox{exp}(iyP)-1]{\tilde \Pi}(dy~dt)\},$$ and the Ito product rule
$$dW_t^2 = dt,\quad \Pi(dy~dt)^2 = \Pi(dy~dt), \eqno(19)$$ with all other
products of stochastic differentials (including $dt$) equal to zero.  Taking
into account (16), one can obtain both backward and forward M.  m. e. (9),
(12) with ${\cal D}={\cal D}^* = C_0^2 ({\bf R})$, the subspace of twice
continuously differential functions with compact support, where the
form-generators correspond to the expression $${\cal L}[X] =
i\beta[P,X]-\frac{\alpha}{2}[P,[P,X]] +\int_{0<|y|}(U_y^* X U_y - X
-iy[P,X]1_h(y))\mu (dy), \eqno(20)$$ defined for $X\in${\sf D}. Here the
first term is the Hamiltonian ``drift'', the second term corresponds to the
interaction with the Gaussian ``white'' noise and is typical for diffusion
approximations, while the last term reflects the influence of the Poisson
``shot'' noises arising in low density limits.  The generator (20) is bounded
if and only if $\alpha=0, \beta=0$ and $\mu(dy)$ is a finite measure on ${\bf
R}\setminus 0$.

The Gaussian noise gives rise to the diffusive generator $${\cal L}[X] =
-\frac{\alpha}{2}[P,[P,X]] = \frac{\alpha}{2}(2PXP - P^2 X - XP^2 ),
\eqno(21)$$ with the obvious standard representation on {\sf D}.  A standard
representation for the last term in (20) can be obtained by taking $$L =
\int_{0<|y|}(I\otimes|y>)(U_y - I)\mu (dy),\quad K = \int_{0<|y|}(I - U_y -
iyP1_h(y))\mu (dy),$$ where $\{ |y>\}$ is the canonical family of ``ket''
vectors in ${\cal H}_0 = L^2 ({\bf R},\mu)$.

From the CCR it follows that the semigroup is covariant with respect to the
representation $y\rightarrow V_y$ describing translations in the momentum
space. As shown in \cite{hol2}, for covariant M. m. e. the non-explosion in
${\cal B}({\cal H})$ is equivalent to the non-explosion in the fixed-point
subalgebra ${\cal A}_V = \{X: V_g^* X V_g = X, g\in G\}$ of the
representation $g\rightarrow V_g$. If this subalgebra is Abelian then the
problem is reduced to the well-studied problem of non-explosion for a
classical Markov process.  In our example the fixed point algebra is the
maximal Abelian subalgebra ${\cal A}_Q$ of operators of the form $X=f(Q)$; by
the CCR $$\Phi_t [f(Q)] = {\sf M}f(Q+\xi_t),$$ and $${\cal L}f (x) =
\beta\frac{df(x)}{dx}+\frac{\alpha}{2}\frac{d^2 f(x)}{dx^2} + \int_{0<|y|}
[f(x+y)-f(x)-yf'(x)1_h(y)]\mu (dy)$$ is the generator of the semigroup
corresponding to the process $\xi_t$ with stationary independent increments,
for which explosion can never occur \cite{feller}.  This is also strictly
related to the additional property of covariance with respect to the space
translations $x\rightarrow U_x$, shared by the semigroup (18). However the
situation is different for more general momentum translation covariant M. m.
e..

To see this, following \cite{hol3}, consider the Hilbert space ${\cal H}= L^2
(l,\infty )$, the domain ${\cal D}=C_0^2 (l,\infty )$ consisting of
continuously twice differentiable functions with compact support, vanishing
at $l$, and the form-generator $${\cal L}(\phi, X, \psi )= <(P+L(Q))\phi |
X(P+L(Q)\psi > -<K\phi | X\psi > - <\phi | XK\psi >, \eqno(22)$$  defined for
$\phi,\psi\in{\cal D} $, where $K=\frac{P^2}{2}+PL(Q)+\frac{|L(Q)|^2}{2}$,
and $L(Q)$ is a continuously differentiable complex function. This
form-generator is covariant with respect to the representation $y\rightarrow
V_y=\mbox{exp}(iyQ)$, and hence the corresponding minimal dynamical semigroup
is also covariant \cite{hol2}. The restriction to the fixed point algebra
${\cal A}_Q$ corresponds to the classical diffusion on $(l,\infty )$ with the
generator $${\cal L}f(x)=2\mbox{Im}L(x)\frac{df(x)}{dx}+\frac{1}{2}\frac{d^2
f(x)}{dx^2}.$$ Non-explosion means that both $l$ and $\infty$ are
non-absorbing boundaries for this diffusion.  The necessary and sufficient
condition for this is Feller's test \cite{ito}, saying that the function
$$\int_{x_0}^x \left[\mbox{exp}\int_x^y  4\mbox{Im}L(z)dz\right] dy,$$ where
$x_0\in (l,\infty )$, must be non-integrable in the neighbourhoods of both
$l$ and $\infty$. In particular, if $ L(x)\equiv 0$ (pure diffusion with no
drift), then the probability of absorption at $l$ is positive, hence the
minimal semigroup is non-unital and the solution of the M. m. e. is not
unique. This minimal semigroup is the extension onto ${\cal B}({\cal H})$ of
the Markov semigroup corresponding to the Brownian motion on $(l, \infty )$
killed at the boundary $l$. Other solutions of the backward M. m. e. are
obtained by taking perturbations corresponding to various boundary conditions
at $l$.  An example of non-standard dynamical semigroup on ${\cal B}({\cal
H})$ is constructed as a singular perturbation of this minimal semigroup
corresponding to rebounding from $l$ to a fixed quantum state $\rho_0$
\cite{hol4}, \cite{hol5}.  \section{Covariant evolutions and the group
cohomology} Consider a backward M. m. e. given by a form-generator ${\cal
L}(\phi, X, \psi)$.  The standard representation (8) of the form-generator is
not unique even if it is subjected to further condition of minimality
\cite{hol3}. If $D$ is a unitary operator in ${\cal H}_0$, $a \in{\cal H}_0$,
and $ b$ is a real number, then the operators $$L' = (I\otimes D)L + I\otimes
a,\quad K' = K + (I\otimes a)^*(I\otimes D)L + [\frac{1}{2}\|a\|^2 - i b
]I,$$ where $(I\otimes a)$ is the operator from $\cal H$ to ${\cal
H}\otimes{\cal H}_0$ acting as $(I\otimes a)\psi = \psi\otimes a$, give
another standard representation for ${\cal L}(\phi, X, \psi)$ satisfying the
minimality condition. The transformations $(D,  a,  b): (L, K)\rightarrow
(L', K')$ form a kind of a ``gauge group'' (cf. also \cite{par}) under the
multiplication law $$(D',  a',  b') (D,  a, b) = (D'D, D' a+ a',  b + b'-
 \mbox{Im}< a'|D' a >).  \eqno(23)$$ We denote this group by $G({\cal L})$.
It is endowed with the natural topology as a subset of the product ${\cal U}
({\cal H}_0 )\times{\cal H}_0\times {\bf R}$, where ${\cal U}({\cal H}_0)$ is
the group of unitary operators in ${\cal H}_0$ with the weak operator
topology.

Let now the form-generator be {\sl covariant} under a (projective) unitary
representation $g\rightarrow V_g$ of a symmetry group $G$, namely, the domain
$\cal D$ be invariant under $V_g$ and $${\cal L}(\phi, V_g^* XV_g, \psi) =
{\cal L}(V_g\phi, X, V_g\psi),\quad \phi, \psi\in{\cal D}.$$

{\bf Theorem 1}. {\sl There is a representation $g\rightarrow (D_g, a_g,
 b_g)$ of $G$ in $G({\cal L})$ such that $$(V_g^* LV_g, V_g^* KV_g) = (D_g,
a_g,  b_g)(L, K).\eqno(24)$$ If $V_g$ is a continuous representation of a
topological group $G$ then the representation $(D_g,  a_g,  b_g)$ is
continuous if and only if the scalar function $g\rightarrow{\cal L}(\phi,
XV_g^*, V_g\psi)$ is continuous for all $\phi, \psi\in{\cal D}, X\in {\cal
B}({\cal H})$.}

The proof of this theorem may be found in \cite{hol3}, and here we discuss
briefly the way it can be applied to find the form of  a covariant generator
for concrete symmetry groups $G$. From this theorem taking into account (24)
it follows that $D_g$ is a unitary representation of $G$ and $ a_g$ is a
first order cocycle for this representation in ${\cal H}_0$: $ a_{g'g} =
D_{g'} a_g +  a_{g'}$. Moreover, the real function $ b_g$ satisfies the
coboundary equation  $ b(g')+ b(g)- b(g'g) = \mbox{Im}< a_{g'}|D_{g'} a_g >$.
Thus the structure of the covariant form-generator is determined by the low
order cohomology of the group $G$, which was studied in detail for many
interesting groups (see, e. g. \cite{cohom}).  In particular, it is well
known that the low order cohomology is trivial for compact groups, that is
every cocycle is a coboundary for such groups, $ a_g = (D_g - I) a$ for some
$ a\in{\cal H}_0$.  It follows that in this case, similarly to the case of
bounded generators, the covariant form-generator has the standard
representation (8) where the c. p. component and the relaxing terms are
separately covariant. That this is not the case for non-compact groups, can
be easily seen from the example of the diffusive generator (21).
\section{Galilean covariant Markovian evolutions} Let $(\xi, \tau )\in {\bf
R}^2$ be a point in the 2-dimensional non-relativistic space-time, and let
$(x, v, t): (\xi, \tau ) \rightarrow (\xi',\tau' ) $ be the Galilei
transformation $$ \xi' = \xi +x+v\tau , \qquad\tau' =\tau +t,   \eqno(25) $$
where $x\in {\bf R}$ is the space shift, $v\in {\bf R}$ the Galilean boost.
For simplicity we consider zero-spin unit mass elementary system
characterized by the Weyl operators $$  W_{x,v} = \mbox{ exp}i(v Q - x P) =
V_v U_x\mbox{exp}(\frac{i}{2}vx),$$ constituting irreducible representation
of the CCR.

A dynamical semigroup $\Phi_t$ is {\sl Galilean covariant} \cite{lanz},
\cite{hol7}, if $$\Phi_t [W_{x,v}^* X W_{x,v}] = W_{x-vt,v}^* \Phi_t [X]
W_{x-vt,v}.$$ Let ${\cal D} \subset\cal H$ be the dense domain $$ {\cal D} =
\bigcap\limits_{x,v\in {\bf R}  } {\rm dom} (v Q - x P)^2,    $$ and let
 $\mbox{\sf D} \subset{\cal B} (\cal H )$ be the domain defined by the
relation (10).  We remark that $\cal D$ is invariant under $W_{x,v}$, and
that {\sf D} is norm-dense in ${\cal T}(\cal H )$ and ${\rm weakly}^*$ -
dense in ${\cal B}(\cal H)$. We make the following assumption

(A)  The domain {\sf D} is contained both in ${\rm dom}{\cal L}_*$ and ${\rm
dom}\cal L$.

{\bf Theorem 2.} {\sl A unital Galilean covariant dynamical semigroup
satisfying the condition (A) has the generator $\cal L$ given by the
following expression on {\sf D} $$ {\cal L}[X] = i[\frac{P^2}{2},X]+
i[{\beta}_P  P + {\beta}_Q  Q, X]$$ $$ - \frac{1}{2} \{ \alpha_{PP} [P,[P ,
X]] + \alpha_{PQ} [P , [Q ,X]] + \alpha_{QQ} [Q ,[Q , X]]\}    \eqno (26)$$
$$ + \int\int_{x^2 + v^2 >0}\{ W_{x, v}^* X W_{x,v} - X - i[xP - vQ, X]1_h
(x,v)\}\nu (dx~dv),   $$ where $\beta_P ,\beta_Q \in {\bf R}$, the real
matrix $$\left[ \begin{array}{cc}\alpha_{PP} & \alpha_{PQ} \\ \alpha_{PQ} &
\alpha_{QQ}\end{array} \right]   $$ is positive definite, $1_h (x,v)$ is the
indicator of the set $x^2 + v^2 \leq h$ and $\nu(dx~dv)$ is a positive
measure on ${\bf R}^2 \backslash \{ 0 \}$ satisfying the Levy condition
$$\int\int_{x^2 + v^2 >0}\{(x^2+v^2)1_h (x,v)+[1-1_h(x,v)]\} \nu (dx~dv) <
\infty. $$ Moreover, the domain {\sf D} is a core for both ${\cal L}_*$ and
$\cal L$ and the corresponding M. m. e. (9), (12) have $\Phi_t$ (resp.
$\Psi_t$) as the unique solution.}

The last statement of the Theorem applies to particular Galilean covariant M.
m. e. arising in various physical applications, such as quantum optics
\cite{car}, precision experiments \cite{brag}, nonlinear quantum mechanics
\cite{doebner} etc.  The uniqueness of the solution of the M. m. e. is
related to the  fact that the fixed point algebra of the representation
$(x,v)\rightarrow W_{x,v}$ is trivial, that is consists of multiples of the
identity operator (cf.  \cite{hol2}).

A derivation of (26) can be based on Theorem 1.  By subtracting from $\cal L$
the Hamiltonian term corresponding to the free motion we obtain a generator
${\cal L}_0$ satisfying the condition of Weyl covariance $$ {\cal L}_0
[W_{x,v}^* XW_{x,v}] = W_{x,v}^* {\cal L}_0 [X] W_{x,v} .$$ Let $L,K$ be the
components of the standard representation of the corresponding
form-generator. According to Theorem 1 there is a unitary representation
$(x,v)\rightarrow D_{x,v}$ of the Abelian group ${\bf R}^2$ and the cocycle $
a_{x,v}$ in the Hilbert space ${\cal H}_0$ such that $L,K$ satisfy the
covariance equations $$W_{x,v}^* LW_{x,v}=(I\otimes D_{x,v})L - I\otimes
a_{-x,-v},$$ $$W_{x,v}^* KW_{x,v}= K - (I\otimes a_{-x,-v})^*L +
[\frac{1}{2}\| a_{x,v}\|^2 - i b_{x,v}]I.$$ These equations can be solved by
diagonalizing the representation $D_{x,v}$ and by using the structure of
cocycles for representations of Abelian locally compact groups \cite{hol6}.
The ``Gaussian'' part of the generator ${\cal L}_0$ arises from the identity
subrepresentation of $D_{x,v}$ while the orthogonal complement gives the
``jump'' part. We conjecture that the assumption (A) can be deduced from the
Galilean covariance itself, as we were able to deduce it from the Weyl
covariance (see \cite{hol7}, where an alternative proof of Theorem 2 is
given).

The generator (20) considered in Section 3  is a particular case of (26),
provided we exclude the free Hamiltonian term.  That generator arose from the
semigroup (18) describing interaction of quantum system with the classical
noise.  It turns out to be possible to give  a similar explicit description
of the Galilean covariant quantum open systems, as systems interacting with
specific classical noises.  Let ${ \xi_t,\eta_t } $ be a classical stochastic
process with stationary independent increments in ${\bf R}^2$, defined by the
characteristic function of the Levy-Khinchin form $$ {\bf M}{\rm exp}i( \mu
 \xi_t - \lambda  \eta_t) = {\rm exp}t\{ i(\mu\beta_P -\lambda\beta_Q)-
\frac{1}{2}(\alpha_{PP} \mu ^2 + 2\alpha_{PQ}\mu  \lambda +
\alpha_{QQ}\lambda ^2 ) $$ $$+ \int\int_{x^2 + v^2 >0} [{\rm e}^{i( \mu  x -
\lambda  v)} - 1-i( \mu  x - \lambda v)1_h(x,v)]\nu (dx~dv)\},  \eqno (27)$$
where $\beta_P ,\beta_Q ; \alpha_{PP} ,\alpha_{PQ} ,\alpha_{QQ}$ and $\nu
(dx~dv)$ are taken from (26). Consider the stochastic differential equations
$$dQ_t = \frac{P_t}{m}dt + d\xi_t,~~~dP_t = d\eta_t,    \eqno (28)   $$ with
the initial conditions $Q_0 = Q, P_0 = P$. These will be the Heisenberg
equations for our open quantum system. They correspond to the infinitesimal
canonical transformation with the Hamiltonian $$dH_t =
\frac{P^2}{2}dt+Pd\xi_t-Qd\eta_t.$$ Defining the chronologically ordered
exponential $$U_t(\xi,\eta)={\cal T}\mbox{exp}(-i\int_0^t dH_s)$$ as the
solution of the corresponding stochastic differential equation, we can prove
(see \cite{hol7}) that $$\Phi_t [X] = {\sf M}U_t(\xi,\eta)^*
XU_t(\xi,\eta).$$ This relation is a generalization of the representation
(18) and the proof proceeds along similar lines by using the stochastic
differential equation for $U_t(\xi, \eta)$ and the distribution of $\xi_t
,\eta_t$ defined by (27).  Equations (28) are the Langevin equation giving
the dilation of the dynamical semigroup $\Phi_t$ with the classical
stationary independent increment processes as the driving noises.
\section{Discussion} The results described in the previous Section are due to
the very restrictive nature of the full Galilean covariance. We obtain much
broader and physically interesting class of quantum Markovian evolutions by
omitting space translations and restricting only to Galilean boosts, that is
to the fundamental symmetry of a non-relativistic particle in a potential
field.  The class of resulting evolutions is described in detail in
\cite{hol7} for the case where the position space is the whole ${\bf R}^3$.
Discussion at the end of Section 3 suggests that contrary to the case of full
Galilean covariance, there is no automatic non-explosion, and boundary
conditions should play an important role, especially for systems with
restricted position domains.  This case deserves much more detailed study.
Other interesting problems are related to introducing spin degrees of freedom
along with spatial ones and to gauge covariance.

Another important distinction of the boost covariant evolutions is that they
describe open systems interacting with quantum rather than classical noises.
This means that the corresponding M. m. e. at least formally can be dilated
to the Langevin equations (see \cite{hol7}) which are quantum stochastic
differential equations driven by quantum Brownian motion or Poisson-type
processes in the sense of \cite{par}, but in general, with unbounded operator
coefficients.  For example, the Langevin equation dilating the diffusive M.
m. e.  defined by the form-generator (22) supplemented with the Hamiltonian
term has the form $$df(Q_t)=i[\frac{P_t^2}{2},f(Q_t)]dt+f'(Q_t)i[(dA_t +
L(Q_t)dt)^{\dagger} - \mbox{h. c.}] + \frac{1}{2}f''(Q_t)dt,$$
$$dP_t=U'(Q_t)dt+i[{\bar L}'(Q_t)(dA_t+\frac{1}{2}L(Q_t)dt) - \mbox{h.c.}],$$
where $U$ is the potential, $A^{\dagger}_t, A_t$ are creation-annihilation
processes representing quantum Brownian motion and h. c. denotes hermitean
conjugated terms.

Remarkably, at least for the minimal solution of M. m. e. there always exists
a representation via solutions of certain {\sl classical} dissipative
stochastic equation in the Hilbert space of the system. It provides a
powerful probabilistic tool for study of the problem of non-explosion for
quantum dynamical semigroups \cite{hol9}, and of the nonlinear stochastic
Schr\"odinger equation arising in the theory of continuous quantum
measurement processes \cite{gat}, \cite{hol9}, \cite{kol}.  \vskip10pt {\sl
Acknowledgements.} The author acknowledges support from Arnold Sommerfeld
Institute for Mathematical Physics, Technical University Clausthal, during
the XXI International Colloquium on Group Theoretical Methods in Physics.
The work was partially supported by RFBR grant no.  96-01-01709.
 \end{document}